\documentclass[journal]{IEEEtran}
\usepackage{amsmath,amsfonts}
\usepackage{algorithmic}
\usepackage{algorithm}
\usepackage{array}
\usepackage[caption=false,font=normalsize,labelfont=sf,textfont=sf]{subfig}
\usepackage{textcomp}
\usepackage{stfloats}
\usepackage{url}
\usepackage{verbatim}
\usepackage{graphicx}
\usepackage{cite}
\usepackage{hyperref}
\usepackage{tabularx}
\usepackage{pifont,booktabs}
\usepackage{makecell}

\usepackage{lipsum}
\usepackage{adjustbox}

\newcommand*{\vertbar}{\rule[-1ex]{0.5pt}{2.5ex}}

\begin{document}

\title{Koopman Regularized Deep Speech Disentanglement for Speaker Verification%
\thanks{This work has been submitted to the IEEE for possible publication.
Copyright may be transferred without notice, after which this version
may no longer be accessible.}}

\author{Nikos Chazaridis,~\IEEEmembership{Student Member,~IEEE}, Mohammad Belal, Rafael Mestre, Timothy J. Norman~\IEEEmembership{Member,~IEEE}, Christine Evers,~\IEEEmembership{Senior Member,~IEEE}
\thanks{N.Chazaridis, Mohammad Belal, Rafael Mestre, Timothy J.Norman and Christine Evers are with the School of Electronics and Computer Science, University of Southampton, SO17 1BJ Southampton, U.K.(e-mail: n.chazaridis@soton.ac.uk; r.mestre@soton.ac.uk; t.j.norman@soton.ac.uk; c.evers@soton.ac.uk)}
\thanks{Mohammad Belal is also with the Marine Geosciences, National Oceanography Centre, SO14 3ZH Southampton, U.K.(e-mail: mob@noc.ac.uk)}}

\markboth{}%
{...}
\maketitle

\begin{abstract}
Human speech contains both linguistic content and speaker dependent characteristics making speaker verification a key technology in identity critical applications. 
Modern deep learning speaker verification systems aim to learn speaker representations that are invariant to semantic content and nuisance factors such as ambient noise.
However, many existing approaches depend on labelled data, textual supervision or large pretrained models as feature extractors, limiting scalability and practical deployment, raising sustainability concerns.
We propose Deep Koopman Speech Disentanglement Autoencoder (DKSD-AE), a structured autoencoder that combines a novel multi-step Koopman operator learning module with instance normalization to disentangle speaker and content dynamics.
Quantitative experiments across multiple datasets demonstrate that DKSD-AE achieves improved or competitive speaker verification performance compared to state-of-the-art baselines while maintaining high content EER, confirming effective disentanglement.
These results are obtained with substantially fewer parameters and without textual supervision. 
Moreover, performance remains stable under increased evaluation scale, highlighting representation robustness and generalization.
Our findings suggest that Koopman-based temporal modelling, when combined with instance normalization, provides an efficient and principled solution for speaker-focused representation learning.
\end{abstract}

\begin{IEEEkeywords}
Speaker verification, disentanglement, Koopman operator theory.
\end{IEEEkeywords}

\section{Introduction}
\IEEEPARstart{S}{peech} signals convey multiple layers of information, including linguistic content, expressive cues, and speaker-specific traits.
Speaker verification (SV) exploits these characteristics to distinguish legitimate users from impostors. 
Effective speaker verification is essential for secure remote services, such as voice-based authentication in financial transactions.
In these applications, systems must achieve high accuracy while also meeting computational efficiency constraints. This motivates learning representations that isolate speaker-specific information from other speech attributes.

Contemporary, deep-learning-based speaker verification systems involve two stages.
First, the enrolment phase, where speaker representations are extracted from target utterances, and next the evaluation phase, where unseen utterances are compared against enrolled speakers..
Classical SV systems leverage speaker representations generated through convolutional neural networks (CNNs), or variants thereof \cite{snyder_x-vectors_2018,desplanques20_ecapa_interspeech}. 
More recently, large pre-trained speech representation models like HuBERT \cite{hsu_2021_hubert} and WavLM \cite{chen_wavlm_2022} have been used as feature extractors for SV systems, either in combination with raw acoustic features or as stand-alone inputs \cite{lian_utts_2023,li_dis_diff_pretrained_2025}.

Existing representation methods, however, require large training corpora.
Classical SV systems require extensive manual annotation, while modern approaches based on large pre-trained speech models depend on thousands of hours of data and substantial computational resources.
While effective, this paradigm is resource-intensive and raises sustainability concerns, particularly when broad, general-purpose models are used for a narrowly defined task such as speaker verification.

Some information sources in speech signals are irrelevant for speaker verification \cite{rose_forensic_speaker_2002}.
Learning representations that retain only speaker-specific characteristics while suppressing irrelevant factors are critical for effective SV.
This observation motivates learning structured representations that explicitly separate speaker-specific information from other speech attributes.
Our aim in this paper, therefore, is to propose a novel speech representation learning method that can effectively decompose speaker identity from other speech attributes without relying on speaker labels or large foundational models, and hence avoid associated annotation and computational costs.

To isolate sources of interest from speech data, we adopt disentangled representation learning.
Disentanglement aims to represent data using separate latent factors that correspond to semantically meaningful sources of variation \cite{wang_disentangled_review_2024}.
In practice, such representations are commonly learned using autoencoders or variational autoencoders (VAEs) \cite{Kingma_vae_2014}.
However, unsupervised disentangled representation learning without additional constraints is fundamentally ill-posed, as the learned representations become highly sensitive to hyperparameter choices and initialization, yielding inconsistent results and poor generalization \cite{locatello_chall_dise_2019}. 
As a consequence, meaningful disentanglement requires the introduction of explicit assumptions, often referred to as inductive biases, that restrict the space of admissible solutions and guide learning.
Despite their appeal, VAE-based disentanglement methods are prone to posterior collapse, as training objectives may over-regularize the latent space while allowing expressive decoders to ignore latent representations. 
This leads to models that fail to disentangle meaningful factors of variation \cite{oord_vqvae_2017}.

Speech signals also exhibit multi-scale temporal structure.
In common with prior work \cite{hsu_unsupervised_2017, yingzhen_DSVAE_2018}, we assume that speaker identity varies more slowly than linguistic content within an utterance. 
Based on this assumption, we model quasi-static speech attributes as slow system dynamics, enforcing this through architectural and operator-theoretic inductive biases. 
Koopman operator theory \cite{Koopman_HamiltonianSpace_1931} offers us a means to achieve this, as it
enables the modelling of system dynamics through a linear, but infinite-dimensional, operator that governs the evolution of latent variables. 
This formulation allows temporal behavior to be explicitly represented and shaped through spectral or structural constraints on the operator \cite{Rowley2009_KMD}.

For SV, linguistic content is a nuisance variable, thus effective speaker representations should be invariant to verbal-content variations.
To explicitly suppress content- and channel-dependent statistics, we incorporate instance normalization \cite{ulyanov_instancenorm_2017} as an inductive bias in conjunction with Koopman operator learning. 
Instance normalization has been shown to facilitate the separation of speaker-dependent characteristics in speech processing tasks such as voice conversion \cite{chou_speech_inorm_2019,chen_again-vc_2021}, but, by itself, it does not guarantee disentanglement of speaker identity and speech content.

In this work we propose a novel autoencoder that disentangles speech spectrograms into speaker and content representations through two dedicated encoder branches.
Temporal separation of the latent space is enforced by combining instance normalization, which captures fast-varying speech content dynamics, with a multi-step Koopman operator learning module designed to model slowly evolving speaker-related attributes.
The multi-step Koopman formulation enables the dynamics encoder to capture long-term temporal structure, thus allowing it to extract fine-grain speaker representations, while the complementary content branch focuses on variability unrelated to identity.
Our contributions are:
\begin{enumerate}
    \item \textbf{Structured disentanglement through temporal inductive bias}. We introduce Deep Koopman Speaker Disentanglement Autoencoder (DKSD-AE), a two-branch architecture that separates fast-varying content dynamics from slowly evolving speaker features.
    Instance normalization promotes invariance to utterance-level variability, while regularized Koopman operator learning models structured temporal evolution of speaker attributes.
    \item \textbf{Multi-step Koopman operator learning for long-term dependency modelling and representation stability}. We propose a novel multi-step prediction formulation that approximates a Koopman operator capable of modelling long-range dynamics in high-dimensional speech data. 
    Ablation studies show that this formulation improves speaker verification performance compared to single-step neural Koopman approaches as well as instance-normalization-only configurations.
    \item \textbf{Accurate, efficient, and scalable speaker verification}. DKSD-AE achieves lower speaker equal error rate (EER) than all baselines on VCTK and all but one on TIMIT, while attaining high content EER that confirms effective disentanglement.
    These results are achieved using significantly fewer parameters compared to baselines, without textual supervision and relying solely on mel-spectrogram inputs.
    Finally, robustness to evaluation scale is demonstrated by only \(\approx1 \)\% EER degradation when increasing test set size from the TIMIT Official test set to the nearly sevenfold larger TIMIT-Full test set, further highlighting representation stability.
\end{enumerate}
The remainder of this article is organized as follows. 
Section~\ref{sec:background} reviews related work to our research. 
Section~\ref{sec:prelims} introduces required theoretical preliminaries for the Koopman operator module.
The DKSD-AE framework is described in Section~\ref{sec:method}.
Experimental setup and results are shown in sections \ref{sec:exp_setup} and \ref{sec:results} respectively. 
We conclude the article in Section~\ref{sec:conclusion}. 
\section{Related Work}
\label{sec:background}
\subsection{Disentangled Speaker Representations}
Sequence-to-sequence autoencoder networks have been largely leveraged for disentanglement due to their architecture allowing the projection of multi-dimensional time-series to a lower dimensional space where features of interest can be isolated.
Recent speech models disentangle utterances into distinct latent spaces representing factors such as speaker identity, linguistic content, and emotion
\cite{hsu_unsupervised_2017,huang_unsupervised_dis_background_2020,lu_speechtriplenet_2023, gengembre_dis_pros_timbre_emot_2024}.

Unsupervised sequence-to-sequence probabilistic methods such as  DSVAE and its extensions \cite{yingzhen_DSVAE_2018,Lian_robust_dsvae_2022,lian_utts_2023} achieve disentanglement by imposing dimensional constraints on two separate latent spaces, with speaker information represented by a fixed-size latent vector and content captured as a sequence of latent variables.
The DSVAE architecture has been extended to leverage contrastive learning by incorporating a mutual information term into the loss function to promote independence between static and dynamic latent variables \cite{Bai_CDSVAE_2021}.
Additionally, several studies utilize multi-branch encoders that partition speech into separate latent representations.
For instance, SpeechTripleNet \cite{lu_speechtriplenet_2023} employs a three-branch VAE where each branch is conditioned on a specific modality to disentangle content, timbre, and prosody.
To guide disentanglement, related approaches incorporate auxiliary modalities alongside raw speech.
VAE-TP \cite{lu_unifying_2024}, for example,  conditions the content branch on textual annotations to enforce the separation of linguistic information from speaker-specific factors. 
While effective, such methods depend on the availability of synchronized metadata during training.
Our model is inspired by these sequence-to-sequence methods that enforce disentanglement through explicit architectural and training biases but without the need for text annotations or speaker labels.
\subsection{Operator Theoretic Representations}
Recently, there has been a surge of methods looking into representation learning of sequential data with operator theoretic representations.
Namely, many approaches leverage the Koopman operator that is understood to linearize a nonlinear system \(\mathbf{F}\) by modelling its transitions with an infinite-dimensional but linear operator \(\mathcal{K}\) \cite{Koopman_HamiltonianSpace_1931}.
The motivation is that a linear surrogate to $\mathbf{F}$ simplifies further analysis compared to the original nonlinear system.
Despite \(\mathcal{K}\) having infinite dimensions, in practice we hope to arrive to a finite dimensional approximation \(\mathbf{K}\) that would largely encode the dynamics in \(\mathbf{F}\).
Dynamic mode decomposition (DMD) \cite{Schmid_dmd_2010} and its variants like the extended DMD (eDMD) \cite{Williams2015_eDMD} are prominent methods of producing an estimate of \(\mathcal{K}\) but can be numerically unstable and sensitive to noise in high-dimensional settings \cite{dawson_characterizing_noise_dmd_2016,Brunton_Koopman_review_2022}.

Nevertheless, estimating  \(\mathcal{K}\) can also be achieved with neural networks namely autoencoders \cite{Takeishi_koopman_ae_2017, Lusch_2018_DeepDynamics}.
Neural Koopman operator representations have been mainly used for forecasting sequential data \cite{Azencot_forecastingDMD_2020,geneva_transformers_koop_2022,Liu_koopa_2023} with less effort made on exploring disentangled representation learning \cite{Berman_SKD_2023, bai_hierarchical_koopman_2025} or speech applications.
In this work, we focus on modelling long-term dynamics in speech sequences and disentangling content and speaker attributes to distinct representations.
Our method is related to the Sequential Koopman Disentanglement (SKD) model \cite{Berman_SKD_2023} that disentangles time-series data into static and dynamic representations by spectrally constraining a \textit{single} Koopman operator \(\mathbf{K}\) that models the full system dynamics.
In contrast to \cite{Berman_SKD_2023}, our approach adopts a structured decomposition.
Koopman operator learning is applied exclusively to the quasi-static speaker-related subspace via a dedicated encoder branch while faster-varying speech components are handled separately with instance normalization and a different encoder branch within the DKSD-AE framework.
Representing multi-scale dynamics of high-dimensional signals with a single matrix as in the SKD places a considerable restriction on the model, especially in light of recent machine-learning approaches that rely on more expressive architectures.
This observation aligns with previous work where authors use Koopman embeddings as a feature extraction step to train a separate predictive model \cite{geneva_transformers_koop_2022} while others employ multiple Koopman operators to capture multi-scale dynamics for disentanglement \cite{bai_hierarchical_koopman_2025} and forecasting \cite{Wang_KNF_2023}.
\section{Preliminaries}
\label{sec:prelims}
\subsection{Koopman Operator Theory}
Many time-varying phenomena can be expressed as a dynamical system by introducing a state \(\mathbf{x}_t\) and a generally nonlinear map \(\mathbf{F}\) on the state space \(\mathcal{X}\):
\begin{equation}\label{dyn_sys3_1}
    \mathbf{x}_{t+1}=\mathbf{F}(\mathbf{x}_t)
\end{equation}
System analysis, in the nonlinear case can be difficult, as certain realizations of nonlinear systems exhibit sensitivity to initial conditions, where small changes in starting values can lead to different outcomes, or lack of global solutions among other tractability issues.
Instead, we can seek a linear surrogate of \(\mathbf{F}\), so that the dynamics can be uncovered with spectral tools available for linear dynamics \cite{Koopman_HamiltonianSpace_1931}.

Koopman operator theory \cite{Koopman_HamiltonianSpace_1931} provides such an alternative by shifting attention from states \(\mathbf{x}_t\) to observables \(g: \mathcal{X} \rightarrow \mathbb{C}\).
The Koopman operator \(\mathcal{K}\) acts linearly on observables by composition.
For a discreet-time system we have:
\begin{equation}
    (\mathcal{K}g)(\mathbf{x}_t) = g(\mathbf{F}(\mathbf{x}_t))=g(\mathbf{x}_{t+1}),
\end{equation}
yielding a linear but infinite-dimensional representation of nonlinear dynamics in a function space \cite{Brunton_Koopman_review_2022}.
Reducing nonlinear dynamics to a linear space allows system analysis by spectral decomposition of the Koopman operator.
However, being infinite dimensional, the Koopman operator presents practical issues for computation and analysis.
To address this limitation, \cite{Takeishi_koopman_ae_2017, Lusch_2018_DeepDynamics} approximate the Koopman operator using finite-dimensional representations from autoencoders. 
By learning nonlinear observable functions through multiple encoder layers, autoencoders generate expressive yet low-dimensional latent representations, in which the Koopman operator can be efficiently estimated via least squares.
\subsection{Autoencoders for Finite Koopman Approximation}
\label{subsec:edmd_ae}
A common data-driven approach to estimate the Koopman operator \(\mathcal{K}\) with a neural network is to design an autoencoder network with a bottleneck that estimates the finite-approximation \(\mathbf{K}\) \cite{Takeishi_koopman_ae_2017}.
In the general Koopman autoencoder setting, the encoder defines a nonlinear lifting of the state \(\mathbf{x}_t\) into a space of observables, while the latent dynamics are constrained to evolve linearly, similarly to the eDMD.
Extended DMD enables the estimation of \(\mathbf{K}\) by modelling the evolution of a finite set of nonlinear observable functions of the state \cite{Williams2015_eDMD}.
In practice, eDMD depends on hand-crafted observable functions, and selecting an appropriate dictionary without prior knowledge of the system dynamics can be difficult.
In contrast, Koopman autoencoders learn these observable functions directly from data.
The resulting output of the encoder \(\mathbf{Z}\) is used to compute \(\mathbf{K}\).

Similar to the eDMD formulation \cite{Williams2015_eDMD,Lusch_2018_DeepDynamics}, common practice in the Koopman autoencoder bottleneck \cite{Takeishi_koopman_ae_2017,Lusch_2018_DeepDynamics}, is to split the latent dynamics tensor \(\mathbf{Z} \in \mathbb{R}^{T\times d}\) into two overlapping snapshots with a difference of \(\Delta t\) (\(\Delta t =1 \) here):
\begin{equation}
\mathbf{Z}^{-} = 
\left[
  \begin{array}{cccc}
    \vertbar & \vertbar &        & \vertbar \\
    \mathbf{z}(t_0)    & \mathbf{z}(t_1)    & \ldots & \mathbf{z}(t_{T-1} )    \\
    \vertbar & \vertbar &        & \vertbar 
  \end{array}
\right],
\end{equation}
\begin{equation}
\mathbf{Z}^{+} = 
\left[
  \begin{array}{cccc}
    \vertbar & \vertbar &        & \vertbar \\
    \mathbf{z}(t_1)    & \mathbf{z}(t_2)    & \ldots & \mathbf{z}({t}_{T})    \\
    \vertbar & \vertbar &        & \vertbar 
  \end{array}
\right]
\end{equation}
Then the best-fit Koopman operator estimate \(\mathbf{K}\) is produced by solving the following regression:
\begin{equation}
\label{k_reg_sing_step}
    \mathbf{K} = \arg\min_{\mathbf{K}} \left\| \mathbf{Z}^{+} - \mathbf{Z}^{-}\mathbf{K}\right\|_{F} = \mathbf{Z}^{+} (\mathbf{Z}^{-})^{\dagger}, 
\end{equation}
where \( \left\|\cdot \right\|_{F}\) denotes the Frobenius norm and \((\mathbf{Z}^-)^{\dagger}\) the Moore-Penrose pseudoinverse of \(\mathbf{Z}^-\).
Commonly, Koopman autoencoders guide learning of \(\mathbf{K}\) by adopting a mean-squared-error loss of the true and predicted latent dynamics tensor \(\mathbf{\widehat{Z}^+}\).
\begin{figure*}
    \centering
    \includegraphics[
        width=\linewidth,
        trim= 25pt 0 100pt 0,
        clip]{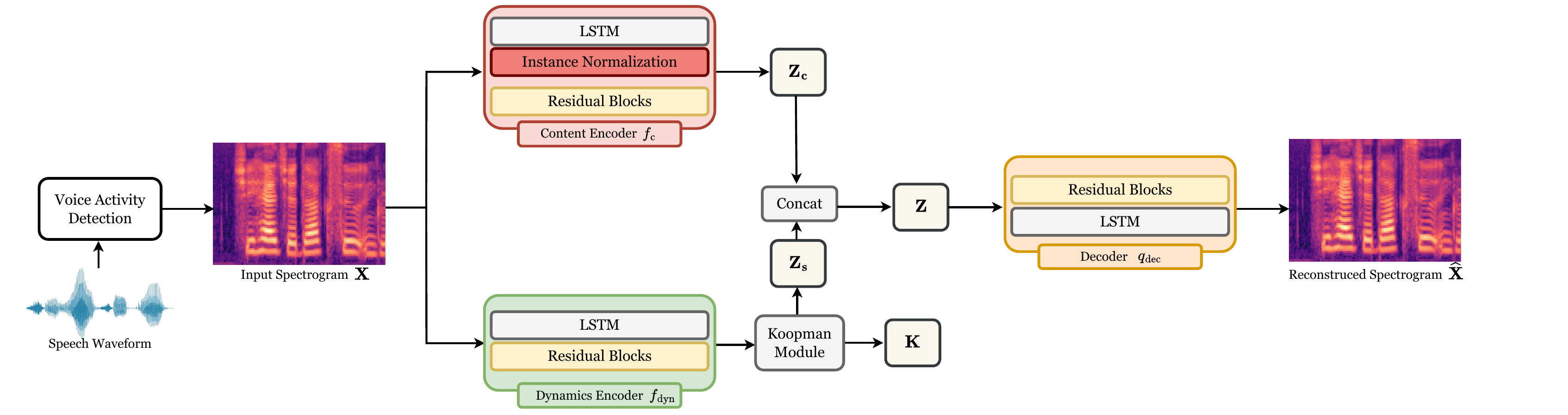}
    \caption{The DKSD-AE framework architecture.
    Speech utterances are processed with VAD and then mel-spectrograms are extracted. 
    Following, the input mel-spectrograms \(\mathbf{X}\) are fed to the dynamics encoder \(f_\text{dyn}\), to learn the Koopman operator \(\mathbf{K}\) and the speaker identity representation \(\mathbf{Z}_s\). 
    Concurrently, \(\mathbf{X}\) is fed to \(f_c\), the content encoder, to learn a content representation \(\mathbf{Z}_c\) via instance normalization.
    Finally, \(\mathbf{Z}_s\) and \(\mathbf{Z}_c\) are concatenated and fed to the decoder, \(q_\text{dec}\), to generate reconstructed mel-spectrograms \(\mathbf{\widehat{X}}\). }
    \label{fig:architecture}
\end{figure*}
\section{Proposed Method}
\label{sec:method}
Our proposed method builds on advances in Koopman autoencoders \cite{Takeishi_koopman_ae_2017,Lusch_2018_DeepDynamics} and instance normalization \cite{chou_speech_inorm_2019} to disentangle speaker and content attributes from speech into distinct representations.
We introduce DKSD-AE, a novel autoencoder neural network architecture with a two-branch encoder and a single decoder \(q_\text{dec}\). 
Let \(\mathbf{X} \in \mathbb{R}^{B\times T\times F}\) denote a three-dimensional input spectrogram data tensor, where \(B\) is the training batch size dimension, \(T\) the number of spectrogram time bins and \(F\) its frequency bins.
The dynamics encoder \(f_{\text{dyn}}\) captures quasi-static attributes from speech into a speaker identity representation \(\mathbf{Z}_\text{s} \in \mathbb{R}^{B\times T\times k}\), by learning a spectrally constrained Koopman operator.
Learning of the Koopman operator \(\mathbf{K}\) is enabled with a novel multi-step forecasting formulation tailored to encode long-term dependencies in input data \(\mathbf{X}\).
Our model learns a content representation \(\mathbf{Z}_\text{c} \in \mathbb{R}^{B\times T\times k}\) from inputs \(\mathbf{X}\) with the content encoder \(f_{\text{c}}\) which captures time-variant attributes that characterize verbal content in speech influenced by instance normalization.
In the following sections, for notation conciseness we will omit the batch dimension when presenting tensors.
Fig. \ref{fig:architecture} illustrates the full DKSD-AE learning pipeline.
\subsection{Koopman Operator Learning for Disentanglement}
\label{sec:koop_module}
To learn a Koopman operator with the dynamics encoder representation \(\mathbf{Z}_{s}\) we follow a different approach than commonly followed in the literature.
It has been shown that estimation of \(\mathbf{K}\) with DMD variants suffers from computational feasibility issues, like overfitting, and the presence of any noise impacts the extraction of meaningful dynamical features from data \cite{dawson_characterizing_noise_dmd_2016,Otto_linearly_recur_dmd_2019,Brunton_Koopman_review_2022}.
To make the computation of \(\mathbf{K}\) robust to noise introduced in the early stages of training, we regularise the Moore-Penrose pseudo-inverse used to compute \(\mathbf{K}\) with an \(\ell_2\)-penalty.
The functional to minimise becomes:
\begin{equation}
\label{k_functional_reg}
    \mathbf{K} = \arg\min_\mathbf{K} \left\| \mathbf{Z}^{+} - \mathbf{Z}^{-}\mathbf{K} \right\|_{F} + \lambda \left\| \mathbf{K}\right\|^2_F, ~~\lambda\geq0
\end{equation}
We solve Eq.~\eqref{k_functional_reg} as following:
\begin{equation}
\label{regularized_K_eq}
    \mathbf{K}=((\mathbf{Z}^{-})^{T}\mathbf{Z}^{-}+\lambda \mathbf{I})^{-1}(\mathbf{Z}^{-})^{T}\mathbf{Z}^{+}
\end{equation} 

\begin{figure}[t!]
    \centering
    \includegraphics[width=\linewidth]{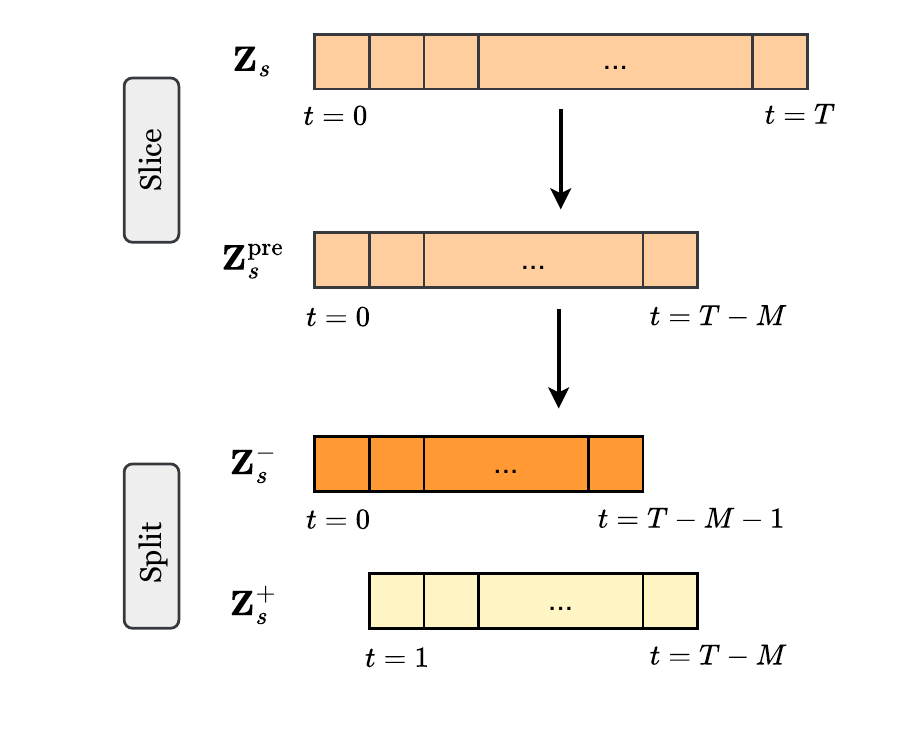}
    \caption{Processing steps to estimate the multi-step prediction loss \(\mathcal{L}_{\text{pred}}\). From top to bottom, first a prefix \( \mathbf{Z}_s^{\mathrm{pre}}\) is extracted from the full dynamics representation \(\mathbf{Z}_s\). 
    Then \( \mathbf{Z}_s^{\mathrm{pre}}\) is split into two subsequent time shifted views, used to solve the the Koopman operator regression shown in Eq.~\eqref{regularized_K_eq}.}
    \label{fig:mult_step_pred}
\end{figure}

Furthermore, unlike common approaches in estimating \(\mathbf{K}\) that split the observables \(\mathbf{Z}\) into two subsequent overlapping windows \cite{Takeishi_koopman_ae_2017,Lusch_2018_DeepDynamics}, we introduce a multi-step forecasting Koopman approximation module designed to capture long-term speech dynamics.
The proposed Koopman module splits \(\mathbf{Z}_{s}\) into \(M\) overlapping windows with a \(\Delta t = 1\). 
We posit that if the estimated \(\mathbf{K}\) can accurately forecast several states in the future, our dynamics encoder has learnt observable functions that model long-term dynamics of speech signals sufficiently.
To that we introduce a loss function, \(\mathcal{L}_\text{pred}\)  that is evaluated on predicting states in the latent observables space with \(\mathbf{K}\) across an \(M\)-step horizon.
Let \(\mathbf{Z}_s=f_{\text{dyn}}(\mathbf{X}) \in \mathbb{R}^{T\times k}\) denote the full dynamics representation where \(k\) is the latent dimension.
To enable multi-step forecasting over a horizon of length \(M\), we first extract a temporal prefix of the sequence,
\begin{equation}
\label{eq:z_prefix}
    \mathbf{Z}_s^{\mathrm{pre}} = \mathbf{Z}_s[0:T-M] \in \mathbb{R}^{(T-M)\times k},
\end{equation}
\begin{equation}
T \in \mathbb{Z}_{>2}, 
\qquad 
M \in \{1,\dots,T-2\},
\end{equation}
to retain a copy of the first \(T-M\) time steps of \(\mathbf{Z}_s\).
Then the Koopman operator is estimated by forming two temporally shifted tensors:
\begin{equation}
\label{eq:split_z_dynamics}
    \mathbf{Z}_s^{-} = \mathbf{Z}_s[0:T-M-1], \quad
    \mathbf{Z}_s^{+} = \mathbf{Z}_s[1:T-M],
\end{equation}
which are used to solve the regularized pseudo-inverse as in Eq.~\eqref{regularized_K_eq}.
The process of slicing \(\mathbf{Z}_s\) across the time dimension and the subsequent extraction of two time-shifted views is illustrated in Fig.~\ref{fig:mult_step_pred}.
Following, \(\mathbf{K}\) is used to forecast latent observable vectors at \(M\) steps in the future.
For each prediction horizon \(m \in \{1, \dots,M \}\) we define the predicted latent dynamics representation as:
\begin{equation}
    \widehat{\mathbf{Z}}_{s}^{(m)} = \mathbf{Z}_s^{-}\mathbf{K}^m \in \mathbb{R}^{(T-M)\times k},
\end{equation}
where \(\mathbf{K}^m\) is \(\mathbf{K}\) to the power of \(m\).
The respective ground truth target tensor becomes:
\begin{equation}
    \mathbf{Z}_s^{(m)} = \mathbf{Z}_s[m:T-M+m-1]
\end{equation}
Learning of the Koopman operator \(\mathbf{K}\) is facilitated by minimizing the mean squared error of the residuals of the predictions:
\begin{equation}
\label{eq:l_pred}
        \mathcal{L}_\text{pred}=\frac{1}{2nM}\sum_i^{n} \sum_{m}^M\left\|\widehat{\mathbf{Z}}_{s,i}^{(m)} - \mathbf{Z}_{s,i}^{(m)} \right\|^{2}_{2},
\end{equation}
Minimizing \(\mathcal{L}_{\text{pred}}\) encourages the dynamics encoder \(f_{\text{dyn}}\) to learn Koopman-observable representations for which the latent dynamics evolve linearly under repeated application of \(\mathbf{K}\).

Complementary to \(\mathcal{L}_{\text{pred}}\) we impose an additional penalty to the Koopman operator, aimed at constraining the spectrum of its eigenvalues to lie close to the unit circle, with emphasis on the neighborhood of the real unit eigenvalue, promoting the modelling of slowly varying dynamics.
Analyzing the spectrum of the eigenvalues of an operator to identify attributes that operate at different temporal rates has been examined before for background modelling \cite{Erichson_dmd_background_2019} and multi-factor disentanglement \cite{Berman_SKD_2023}.
Specifically, it is well known that features of a signal that change slowly in time correspond to discreet-time eigenvalues \(\lambda\) of the Koopman operator that satisfy \(|\lambda| \approx1\) \cite{Erichson_dmd_background_2019}.

We revisit the same notion of \cite{Erichson_dmd_background_2019} to constrain the Koopman operator in encoding speech attributes that vary slowly or are static in time.
To encourage stable dynamics in the learned Koopman operator, we adopt the eigenvalue penalty introduced in the SKD method \cite{Berman_SKD_2023}.
While training, after estimating \(\mathbf{K}\), we compute its eigen-decomposition and use the eigenvalues \(\lambda_i\) to calculate the eigenvalue loss \(\mathcal{L}_{\text{eigen}}\) as following:
\begin{equation}
    \mathcal{L}_{\text{eigen}} = \frac{1}{k}\sum_{i}^{k}|\lambda_i - (1+0j)|^2.
\end{equation}
In the SKD formulation, this penalty is one of two disentanglement constraints imposed on 
\(\mathbf{K}\), an operator learnt from whole system dynamics, whereas in our model it is the sole constraint applied to the eigenvalue spectrum of \(\mathbf{K}\), estimated on quasi-static speech dynamics.
This design choice reflects the fact that \(f_\text{dyn}\) exclusively encodes quasi-static speech characteristics, making a single eigenvalue constraint effective in practice.
The eigenvalue loss \(\mathcal{L}_{\text{eigen}}\) is designed to penalize deviations of the Koopman eigenvalues from the unit circle, thus allowing it to capture static and slow-varying speech features dominated by speaker characteristics.
\subsection{Dynamics Encoder}
The proposed dynamics encoder \(f_{\text{dyn}}\) contains a stack of LSTM \cite{hochreiter_LSTM_1997} blocks followed by skip-connection blocks \cite{he_resnets_2016} that reduce the dimensionality of the generated dynamics representation.
We use skip connection blocks as they have been shown to improve the performance of speaker representations for SV systems \cite{li_resnet_speech_2017}. 
Residual or skip connection blocks allow the expansion of the network with additional layers without degrading its performance by estimating a residuals function:
\begin{equation}
    f_{\text{skip}}= \mathbf{Z}_{\text{hid}} + h(\mathbf{Z}_\text{hid}),
\end{equation}
where \(h(.)\) is an arbitrary activation function and \(\mathbf{Z}_{\text{hid}}\) the output of the previous layer.
In DKSD-AE \(h(.)\) is the hyperbolic tangent function.
Thus, we learn iteratively more fine-grained representations with each additional block and avoid computational instability issues while training.
The dynamics encoder of DKSD-AE computes latent observables that are subsequently used as input to the multi-step forecasting Koopman module and respective Koopman loss functions \(\mathcal{L}_\text{pred}\) and  \(\mathcal{L}_\text{eigen}\).
\subsection{Content Encoder}
In preliminary experiments, we investigated whether a single-operator Koopman formulation as in the SKD framework \cite{Berman_SKD_2023} could achieve stable disentanglement.
We reimplemented the architecture described in SKD and evaluated it under the same preprocessing pipeline, training protocol, and evaluation metrics adopted for DKSD-AE.
While SKD is effective for image sequences, under our speech-based experimental setting we could not replicate their EER results.
This suggests that constraining the eigenvalue spectrum of a single Koopman operator may not fully capture the multi-scale structure present in high dimensional speech signals.
Our observation is consistent with recent studies suggesting that modelling multi-scale nonlinear dynamics may require more expressive or structured operator formulations \cite{geneva_transformers_koop_2022,bai_hierarchical_koopman_2025}.
The above findings motivated the introduction of additional inductive structure in our model design.

To promote the separation of slow and faster temporal scales, we incorporate an additional encoder with instance normalization layers.
Empirically, we observe that instance normalization promotes the decoupling of latent dynamics in speech utterances, encouraging fast-varying speech attributes to be captured by our content encoder while allowing \(f_{\text{dyn}}\) to extract fine-grained speaker representations.
The content encoder \(f_{\text{c}}\) is a stack of LSTM layers \cite{hochreiter_LSTM_1997} combined with instance normalization \cite{chou_speech_inorm_2019}, followed by skip-connection blocks.
Since the introduction of instance normalization in \cite{chou_speech_inorm_2019} several different realizations have been proposed that normalize representations across the time axis or frequency axis \cite{lertpetchpun_tinorm_2023}.
We use instance normalization blocks that compute the mean and standard deviation of the input spectrogram across the frequency dimension to normalize out global per-utterance statistics strongly correlated with channel and speaker attributes, as following:
\begin{equation}
\begin{aligned}
\boldsymbol{\mu}_{F} &= \frac{1}{F} \sum_{i=1}^{F} \textbf{Z}_{f,:}, & 
\boldsymbol{\sigma}_{F} &= \sqrt{\frac{1}{F}\sum_{f=1}^{F} \left(\textbf{Z}_{f,:} - \boldsymbol{\mu}_F\right)^{2}+\epsilon},
\end{aligned}
\end{equation}
where \(\mathbf{Z}_{f,:}\) denotes all time indices at frequency bin \(f\).
Then each instance normalization layer block outputs the following embedding:
\begin{equation}
    \textbf{Z}_{\text{IN}} = \frac{\textbf{Z}-\boldsymbol{\mu}_F}{\boldsymbol{\sigma}_F}
\end{equation}
Different to the instance normalization blocks of Chou et al. \cite{chou_speech_inorm_2019} our formulation first processes spectrograms with LSTM blocks as shown in Fig~\ref{fig:architecture}.
The impact of instance normalization is two-fold in the DKSD-AE that is comprised of two encoders. 
It allows the content encoder \(f_{\text{c}}\) to capture speech content information by removing channel and speaker attributes. 
At the same time it forces the dynamics encoder \(f_{\text{dyn}}\) to learn the discarded speaker attributes such that the decoder \(q_{dec}\) can sufficiently reconstruct the input spectrogram \(\mathbf{X}\).
\subsection{Decoder}
The decoder network \(q_\text{dec}\) of our proposed architecture reconstructs the input spectrograms \(\mathbf{X}\) by processing the concatenation of \(\mathbf{Z}_c\) and \(\mathbf{Z}_s\) across the latent dimension:
\begin{equation}
    \mathbf{Z}=
\operatorname{concat}_{\text{feat}}\!\left({\mathbf{Z}}_s, \mathbf{Z}_c\right),
\end{equation}
It is a stack of residual blocks that are upscaling the feature dimension of \(\mathbf{Z}\), followed by LSTM layers.
Finally, a linear layer projects the LSTM hidden states to the frequency bins dimension of the input data.
Learning is driven by a mean squared error of the reconstructed spectrograms as following:
\begin{equation}
    \mathcal{L}_\text{rec} = \frac{1}{N}\sum_{i=1}^N \left\| \widehat{\mathbf{X}}_i-\mathbf{X}_i\right\|_2^2.
\end{equation}

The total loss of our network is presented below:
\begin{equation}
    \mathcal{L}_\text{total} = w_\text{rec}\mathcal{L}_\text{rec} + w_\text{pred}\mathcal{L}_\text{pred}+w_\text{eigen}\mathcal{L}_{\text{eigen}},
\end{equation}
where \(w_\text{*}\) are tunable hyperparameters weighting the total loss to ensure consistent scale of values from the individual terms.

\subsection{Masked Augmentation to Capture Intra-Speaker Variation}
Disentanglement for speaker verification requires both separation of content and identity attributes as well as adequate representation of within-speaker variability. 
The dynamics encoder \(f_\text{dyn}\) is designed to capture fine-grained speaker identity cues so that same speaker representations remain compact in latent space but also separated from those of different speakers.
To further enable encoding of intra-speaker variation we use SpecAugment time and frequency masked augmentations of input spectrograms \cite{park_specaugment_2019}. 
SpecAugment applies random masking on adjacent regions along the time and frequency dimensions of speech spectrograms to generate augmented views during training.
These augmented views expose the model to perturbations that preserve speaker identity, encouraging the encoder to focus on consistent speaker-specific characteristics.

Further, as augmentation increases the difficulty of reconstructing spectrograms, our model is first pretrained optimizing only for the reconstruction loss \(\mathcal{L}_\text{rec}\) for a fixed number of epochs.
Pretraining details are provided in the following section.
\section{Experimental Setup}
\label{sec:exp_setup}
\begin{table}[!t]
    \begin{center}
        \caption{DKSD-AE architecture number of layers and hidden node dimensions.}
        \label{tab:DKSD_AE_LAYERS}
        \begin{tabular}{c c c}
            \toprule
            \textbf{Component} & \textbf{LSTM Blocks} & \textbf{Residual Blocks} \\ [1ex]
            \hline
            \(f_\text{c}\) & 256,128,128,64 & 64,64 \\
            \(f_\text{dyn}\) & 256,128 & 128,128,64,64,64,64,64 \\
            \(g_\text{dec}\) & 128 & 64,64,128  \\
            \bottomrule
        \end{tabular}
    \end{center}
\end{table}
\subsection{Training}
During training, batches of speech spectrograms \(\textbf{X}\) are fed to both encoder branches to generate dynamics representations \(\textbf{Z}_s\) and speech content representations \(\textbf{Z}_c\). 
Our models are trained with early-stopping for 500 epochs and we pretrain for the first 30 epochs optimizing only for reconstructions loss \(\mathcal{L}_{\text{rec}}\).
After pretraining, the model is optimized for the full loss \(\mathcal{L}_\text{total}\).
During full training, \(\textbf{Z}_s\) is used to compute the Koopman operator estimate \(\textbf{K}\) with the Koopman module, as described in Section~\ref{sec:koop_module}.
Following, the concatenation of \(\textbf{Z}_s\) and \(\textbf{Z}_c\) are fed to the decoder \(q_\text{dec}\) to generate reconstructions \(\widehat{\textbf{X}}\).
We use SpecAugment transforms \cite{park_specaugment_2019} to augment the data with \(0.5\) probability of an utterance being altered by an augmentation function.
The individual loss weights identified through experimentation are \(w_\text{rec}=1\), \(w_\text{pred}=0.1\) and \(w_\text{eigen}=5\).

The architecture details of our proposed method are presented in Table~\ref{tab:DKSD_AE_LAYERS}.  
The resulting dynamics representation from \(f_{\text{dyn}}\) is of shape \(\textbf{Z}_s \in \mathbb{R}^{T\times B \times k}\) with \(k=64\). 
The estimated \(\textbf{K}\) has shape \(k \times k\).
Regarding the the content encoder \(f_{c}\) of LSTM-IN blocks, we note that the output of each LSTM is immediately passed through an instance normalization block before being fed to the next layer. 
Additionally, in the decoder \(q_\text{dec}\), the residual blocks precede the LSTM block.

Our models use the AdamW optimizer \cite{loshchilov_adamw_2018} with a learning rate of \(10^{-4}\) and a weight decay of  \(\alpha=0.4\).
We use hyperbolic tangent activation functions for both the LSTMs and the linear layers of the residual blocks.
Finally, our Koopman module splits the input dynamic embedding \(\mathbf{Z}_s\) into \(M=5\) overlapping sliding windows with a \(\Delta t=1\).
The Koopman operator is computed from windows \(m_1\) and \(m_2\) and \(\mathcal{L}_\text{pred}\) is evaluated on forecasting \(M\) steps ahead. 
\subsection{Data}
We evaluated our proposed method on the VCTK \cite{yamagishi_cstr_2019} and TIMIT \cite{garofolo_john_s_timit_1993} datasets. 
VCTK contains 110 speakers with about 400 utterances each.
TIMIT includes a total of 630 speakers with 10 utterances per speaker.
To facilitate one-to-one comparison with existing benchmarks for VCTK, we use 80\% /10\%/10\% split for training, validation and testing as well as downsample audio to 16 kHz. 
Our splits are stratified based on speaker gender.
For TIMIT we follow the official split, that is, 462 speakers for training and 24 speakers for testing, but also present results on the full test set containing 167 speakers.
Both datasets are pre-processed with a custom WebRTC-based \cite{webrtc_vad} voice activity detector (VAD) pipeline to extract voice-only segments from speech. 
We extracted 80-bin log-mel spectrograms with a window length of 50 ms and a hop length of 12.5 ms, as input \(\mathbf{X}\) to the model.
\subsection{Metrics}
We evaluate speaker verification performance with the equal error rate (EER), defined as the operating point at which the false acceptance rate (FAR) equals the false rejection rate (FRR).
Assuming that \(\mathbf{Z}_s\) encodes speaker specific information and \(\mathbf{Z}_c\) encodes linguistic content, effective disentanglement is indicated by consistently low EER when using \(\mathbf{Z}_s\) for SV and near-chance or substantially degraded EER when using \(\mathbf{Z}_c\), reflecting the absence of speaker information in the content representation.
FAR and FRR are computed by first extracting \(\mathbf{Z}_s\) and \(\mathbf{Z}_c\) from  utterances. 
Following, we calculate the cosine similarity scores for same-speaker pairs and different-speaker pairs.
A decision threshold $\epsilon \in [-1,1]$ is then swept over the similarity scores.
A trial is classified as “accept” if its similarity is $\geq \epsilon$ and “reject” otherwise.
For each $\epsilon$, FAR and FRR are computed as below:
\begin{align}
    \text{FAR}(\epsilon)&=\frac{\# \text{imposter trials accepted at}\;\epsilon}{\# \text{total imposter trials}},\\
    \text{FRR}(\epsilon)&=\frac{\# \text{genuine trials rejected at}\;\epsilon}{\#\text{total genuine trials}},
\end{align}
\section{Results: Speaker Verification}
\label{sec:results}
\begin{table*}[!t]
\centering
\caption{Best run performance of proposed approach against the VCTK baseline models in speaker verification. 
DKSD-AE performs better in the SV task than competing approaches without the need for text-annotations or additional modalities. 
Best performance in each EER category is in bold.}
\label{tab:EER_VCTK}
\newcolumntype{Y}{>{\centering\arraybackslash}X}
\begin{tabularx}{\textwidth}{l c c Y c}
    \toprule
    \textbf{Method} & \textbf{EER Speaker} $\downarrow$ & \textbf{EER Content} $\uparrow$ & \textbf{Modality Type} & \textbf{Parameters} \\
    \midrule
    DKSD-AE (Ours) & $\mathbf{2.77\,\%}$ & $\underline{44.0\,\%}$ & Mel Spectrograms & 3.5M \\
    SpeechTripleNet \cite{lu_speechtriplenet_2023} & $7.01\,\%$ & $\mathbf{45.5}\,\%$ & Mel Spectrograms, Pitch and Energy Contours & 19.3M \\
    VAE-TP \cite{lu_unifying_2024} & $2.90\,\%$ & $44.1\,\%$ & Mel spectrograms and Text Annotations & 399M \\
    UTTS \cite{lian_utts_2023} & $3.90\,\%$ & $43.2\,\%$  & WavLM\cite{chen_wavlm_2022} Features and Text Annotations & \(>\)94M \\
    \bottomrule
\end{tabularx}
\end{table*}
\subsection{Robust Speaker Verification with Parameter Efficiency}
We compare DKSD-AE with state-of-the-art disentanglement models for SV using EER as the evaluation metric.
Across datasets, DKSD-AE achieves competitive or improved SV performance while requiring substantially fewer parameters.
Tables \ref{tab:EER_VCTK} and \ref{tab:EER_TIMIT} compare our best-performing models against baselines on the VCTK and TIMIT datasets, respectively, while Table \ref{tab:EER_diff_seeds} reports results across five random seeds on VCTK, TIMIT Official, and TIMIT-full, including mean and standard deviation to assess robustness.
\begin{table}[!b]
    \begin{center}
        \caption{Our model performs on par or better than state-of-the-art approaches, with signficant improvements in the parameter footprint compared to VAE variants.}
        \label{tab:EER_TIMIT}
        \begin{tabularx}{\columnwidth}{l c c c}
            \hline
             \textbf{Method} & \textbf{EER Speaker} $\downarrow$ & \textbf{EER Content} $\uparrow$ & \textbf{Parameters}\\
            \hline
            DKSD-AE (Ours) & \underline{3.90} \% & \(\mathbf{45.8}\) \% & 3.5M \\
            DSVAE \cite{yingzhen_DSVAE_2018} & 4.78 \% & 17.84 \% & 21M \\
            C-DSVAE \cite{Bai_CDSVAE_2021} & 4.03 \% & 31.81 \% & 11M \\
            SKD \cite{Berman_SKD_2023} & 4.46 \% & 26.78 \% & 2M \\
            D-DSVAE \cite{Lian_robust_dsvae_2022} & \(\mathbf{3.25}\) \% & 38.83 \% & \textit{ND}\\
            \hline
        \end{tabularx}
    \end{center}
\end{table}

On VCTK, we evaluate against disentanglement autoencoders conditioned on distinct inductive biases.
Specifically, SpeechTripleNet \cite{lu_speechtriplenet_2023} which combines mel-spectrograms with pitch and energy features, VAE-TP \cite{lu_unifying_2024} that leverages text annotations to learn a text-conditioned prior that guides the content representation and UTTS \cite{lian_utts_2023} which relies on self-supervised acoustic features extracted from the large pretrained WavLM model \cite{chen_wavlm_2022} in addition to textual supervision.
For TIMIT, we present results against the original DSVAE and its extensions, namely the C-DSVAE, D-DSVAE and SKD models that are more thoroughly presented in Section~\ref{sec:background}.

Tables~\ref{tab:EER_VCTK} and~\ref{tab:EER_TIMIT} show that DKSD-AE achieves lower speaker EER than all baseline methods on the VCTK dataset and all except for one on the TIMIT dataset.
Moreover, this performance is achieved with substantially fewer trainable parameters and without relying on additional modalities beyond mel spectrograms, highlighting the impact of our design in efficient SV.
The consistent trends across VCTK and TIMIT suggest that the proposed inductive biases generalize across datasets with different recording conditions.
Our model also outperforms all but one baselines in content EER, further validating disentanglement as \(\mathbf{Z}_c\) is uninformative for speaker verification.

Additionally, Table~\ref{tab:EER_diff_seeds} reports speaker and content EER on VCTK, TIMIT, and the substantially larger TIMIT-full test set, reporting mean and standard deviation across different seeds. 
Interestingly, \textbf{despite the nearly sevenfold increase in evaluation scale from TIMIT Official test set to the TIMIT-Full test set, speaker EER degrades by only \(\approx\) 1\%}, indicating that the proposed representations generalize well and remain stable as the number of speakers increases.
Reporting mean EER performance and respective standard deviation across five different seeds further demonstrates the stability of the proposed approach, as the observed standard deviation remains consistently low across datasets, indicating limited sensitivity to random initialization.

\begin{figure}[!b]
    \centering
    \includegraphics[width=0.95\linewidth]{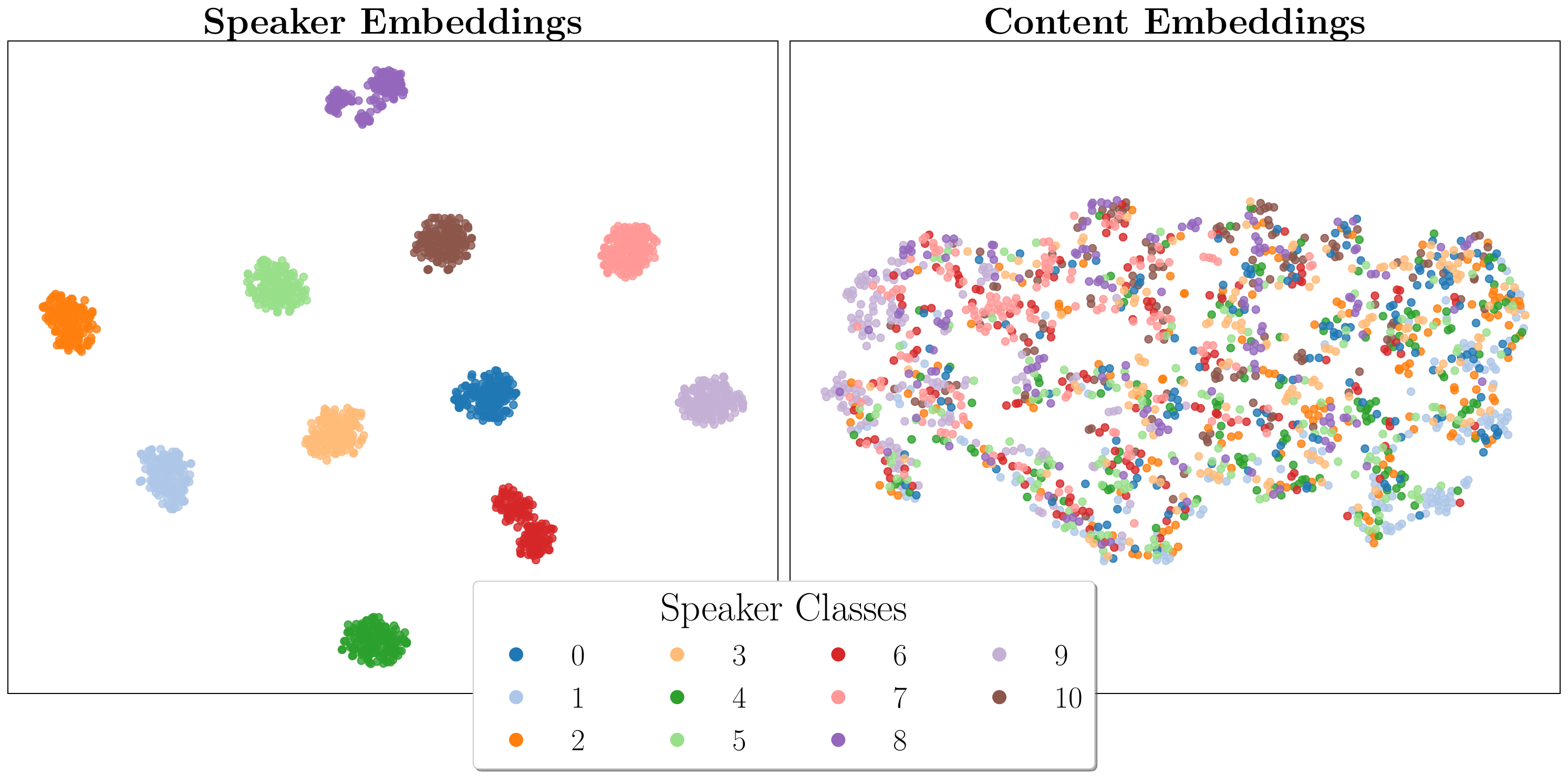}
    \caption{Visual comparison of speaker \(\mathbf{Z}_s\) (left) and content representations \(\mathbf{Z}_c\) (right), after dimensionality reduction with PCA and t-SNE \cite{maaten_t_sne_2008}.
    Same speaker representations \(\mathbf{Z}_s\) (left) form compact well-separated clusters in 2D, whereas content representations \(\mathbf{Z}_c\) are dispersed without speaker-specific grouping (right).
    Different colours correspond to different speaker classes.}
    \label{fig:content_vs_speaker_2d}
\end{figure}

Complementary to the EER results, we validate the effectiveness of the proposed disentaglement approach through visualizations of the learnt disentangled representations.
Speaker and content representations are extracted from test speaker utterances and are projected in two dimensions with PCA and via t-SNE \cite{maaten_t_sne_2008} for comparison.
Fig.~\ref{fig:content_vs_speaker_2d} illustrates that speaker representations \(\mathbf{Z}_s\) are forming well defined clusters for each speaker class whereas content embeddings \(\mathbf{Z}_c\) exhibit no speaker specific structure and are scattered across the two-dimensional plane, indicating successful disentanglement.

Our method relying only on speech data allows DKSD-AE to be more flexible in that it can be applied to different speech datasets with minimal tweaking.
In contrast, without access to text data, methods requiring both audio and text modalities are limited to datasets with aligned transcriptions and may not be directly applicable otherwise. 
This is apparent in the case of UTTS \cite{lian_utts_2023}, where authors exclude speaker ``p315" from their experiments as their transcriptions are not available in the VCTK corpus.
By avoiding the need of text labels, our method is also computationally more lightweight than the compared approaches, allowing deployment on resource constrained devices.
\begin{table}[!t]
    \begin{center}
        \caption{DKSD-AE shows consistent EER performance across VCTK, TIMIT and TIMIT-TEST on different random initializations.}
        \label{tab:EER_diff_seeds}
        \begin{tabular}{c c c}
            \toprule
            \textbf{Dataset} & \textbf{EER Speaker} $\downarrow$ & \textbf{EER Content} $\uparrow$\\
            \hline
             VCTK & \({3.44} \pm 0.56\) \% & \({44.70} \pm 1.50\)  \% \\

            TIMIT Official  & \(4.68\pm0.69\) \% & \(46.04\pm0.59\) \% \\

            TIMIT Full  & \(5.64\pm0.27\) \% & \(45.94\pm0.62\) \% \\
            \bottomrule
        \end{tabular}
    \end{center}
\end{table}
\subsection{Regularized Operator Theoretic Representations Enhance Speech Disentanglement}
We demonstrate the enhanced disentanglement performance from learning operator theoretic representations with our proposed \(\mathcal{L}_\text{total}\) by comparing different realizations of the Koopman module for disentanglement.
Namely, we compare our proposed model to a version without the eigenvalue penalty \(\mathcal{L}_{\text{eig}}\), which simply learns a Koopman operator with \(\mathcal{L}_{\text{pred}}\), and to a version without Koopman learning.
Table~\ref{tab:koop_ablation_on_off} highlights that training with the full objective \(\mathcal{L}_\text{total}\) yields the best speaker and content EER performance.
Our results showcase that instance normalization alone enables the content encoder \(f_\text{c}\) to capture speech content and thus forces \(f_\text{dyn}\) to encode high-level speaker attributes.
However, \textbf{conditioning} \(f_\text{dyn}\) \textbf{to learn Koopman observable functions with multi-step temporal consistency allows learning finer-grained speaker features} and reduces slow feature leakage into content representations \(\mathbf{Z}_c\) compared to the instance-normalization-only variant.

Speech carries features evolving at different temporal scales.
We enforce a hierarchical disentanglement, by modelling fast-varying attributes with instance normalization and slowly varying speaker dynamics via multi-step Koopman prediction in \(f_\text{dyn}\).
Expecting a single linear operator to efficiently model the propagation of high-dimensional, multi-scale speech features presents significant limitations, as noted in the recent literature \cite{geneva_transformers_koop_2022, bai_hierarchical_koopman_2025}.
Unlike prior work in disentangling speech sequences with Koopman embeddings \cite{Berman_SKD_2023}, we do not model the system using a single operator.
\begin{table}[!t]
    \centering
    \caption{The complete objective yields improved SV performance and stronger disentanglement compared to both a reconstruction-only baseline and a Koopman autoencoder trained solely with \(\mathcal{L}_\text{pred}\) across different datasets and different seeds.}
    \label{tab:koop_ablation_on_off}
    \resizebox{\columnwidth}{!}{
        \begin{tabular}{c c c c}
            \toprule
            Dataset & Koopman Learning & EER Speaker & EER Content \\
            \hline
              &  \(\mathcal{L}_{\text{total}}\) & \(\mathbf{3.66}\pm {0.55}\) \%  &\( \mathbf{44.36}\pm0.93\)\% \\
             VCTK & \(\mathcal{L}_{\text{pred}}\)  &  \(5.27\pm0.49\) \%  & \( 35.04\pm3.97\)\%\\
              & \(\mathcal{L}_\text{rec}\) &  \(4.65\pm0.25\) \% & \( 39.04\pm1.95\)\% \\
            \hline
             & \(\mathcal{L}_\text{total}\) & \(\mathbf{4.68} \pm 0.69\) \% & \(\mathbf{46.04}\pm0.59\) \% \\
             TIMIT & \(\mathcal{L}_\text{pred}\) & \(7.15\pm0.50\) \%& \(39.82\pm1.98\) \%\\
              & \(\mathcal{L}_\text{rec}\) & \(6.47\pm1.18\) \%& \(43.39\pm3.38\) \%\\
            \bottomrule
        \end{tabular}
        }
\end{table}
\subsection{Multi-step Koopman Loss Enables Long-term Dynamics Capturing in Speech}
\begin{figure}[!t]
    \centering
    \includegraphics[width=\linewidth]{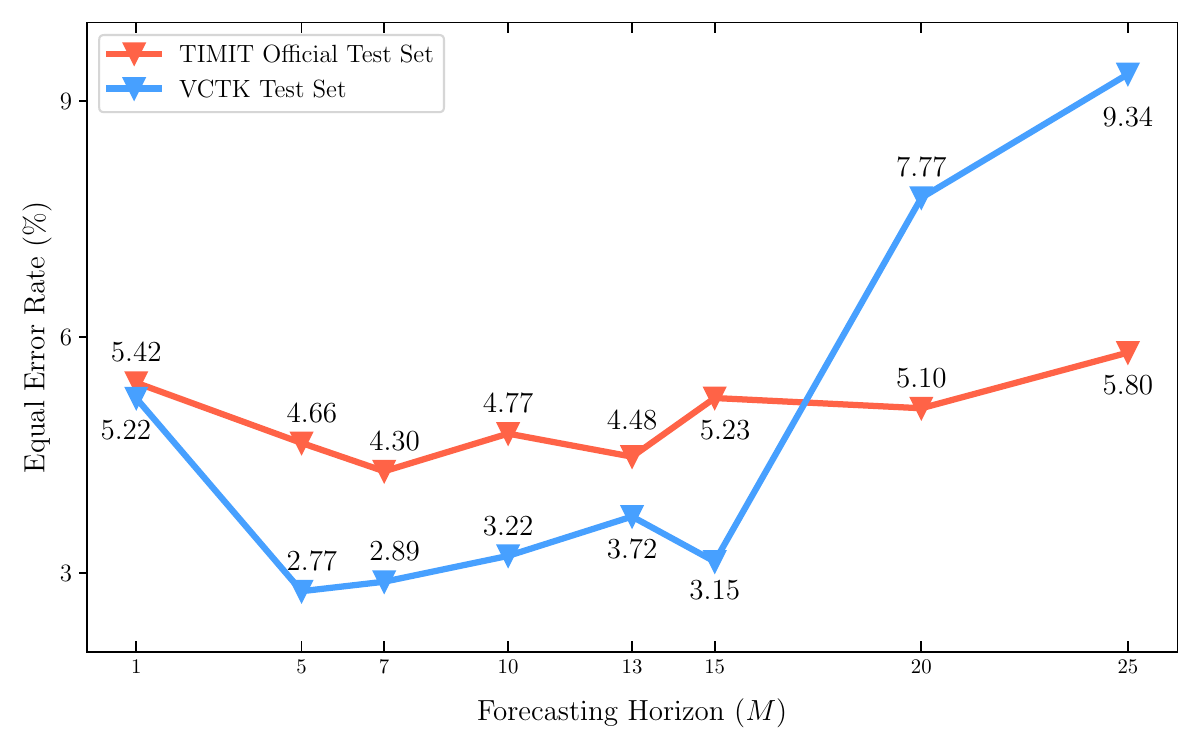}
    \caption{Our multi-step Koopman operator learning module formulation presents improvements in speaker EER when the forecasting horizon \( M \in [5,15]\) for VCTK and when \( M \in [5,13]\) for TIMIT although less pronounced. 
    The identified effective \( M\)-range  corresponds to forecasting horizon length between \(5 \%\) and \(10\%\) of the initial speech utterance duration. }
    \label{fig:m_ablation}
\end{figure}

In this section, we show that the proposed multi-step prediction loss \(\mathcal{L}_\text{pred}\) enables the Koopman operator to capture long-term dependencies in speech.
This is demonstrated through an ablation study, where multi-step training improves speaker EER compared to single-step \(\mathcal{L}_\text{pred}\) formulations across TIMIT and VCTK.
The ablation results are presented in Fig.~\ref{fig:m_ablation}.
We present the EER of speaker embeddings \(\mathbf{Z}_s\) from models with the same configuration that differ only on the choice of \(M\).
Multi-step \(\mathcal{L}_\text{pred}\) models consistently outperform the Koopman autoencoder formulation where \(M=1\), in disentanglement with the EER, up to a forecasting horizon of \(M=15\) steps ahead for the VCTK dataset and \(M=13\) for TIMIT.
When approximating \(\mathbf{K}\) with \(M=1\) the Koopman observable functions estimating the operator need only to capture one-step-ahead dynamics to ensure convergence of a single-step \(\mathcal{L}_\text{pred}\), which may resort to overfitting the encoder on that limited horizon, ignoring past or future dynamics.
In contrast, while we learn \(\mathbf{K}\) from the first two sliding windows of \(\mathbf{Z}_s^{\mathrm{pre}}\), as presented in Eq.~\eqref{eq:split_z_dynamics}, our \(\mathcal{L}_\text{pred}\) is evaluated \(M\) states ahead, enabling \(f_\text{dyn}\) to learn observable functions that capture longer-term dynamics.

Multi-step temporal consistency of \(\mathbf{K}\) acts as a powerful regulariser for disentanglement, regardless of the underlying dynamical parameterisation.
The benefits of multi-step \(\mathcal{L}_\text{pred}\) are consistent across datasets, although  less pronounced on TIMIT, possibly due to its higher number of speakers compared to VCTK.
Fig.~\ref{fig:m_ablation} also shows an effective \(M\)-range where the system dynamics can be efficiently linearized with Koopman observables and thus enable good performance in downstream representation learning tasks.
When \(M\) exceeds the identified range, SV performance degrades as multi-step prediction becomes increasingly challenging in the high-dimensional latent space.
\section{Conclusion}
\label{sec:conclusion}
This paper presented a deep autoencoder to disentangle speech in speaker identity and verbal content representations for speaker verification.
We introduced DKSD-AE which combines multi-step Koopman operator learning and instance normalization as inductive biases to encourage separation of fast-varying content dynamics from slowly evolving speaker-related attributes.
Experimental evaluation yields speaker representations that are discriminative for verification and invariant to linguistic content.
This is reflected by competitive or improved SV performance, alongside high content EER against baselines, achieved with reduced parameter count and without reliance on textual supervision.
The proposed approach remains robust under increased evaluation scale and across multiple datasets, highlighting its generalization capability.
Overall, our findings show that structured Koopman operator learning in conjunction with instance normalization provides an efficient and principled avenue to disentangle speaker and speech content attributes.

The current evaluation is constrained to text-independent speaker verification, and extending it to emotional speech or acoustically degraded conditions remains future work.
A natural next step is to couple the proposed Koopman-based dynamics module with transformer-style encoders \cite{latif_speech_transformers_2025}, leveraging their strong sequence modelling capability to scale to longer and more diverse utterances while preserving the proposed inductive biases.
\section*{Acknowledgments}
\noindent This work was supported by the UK Research and Innovation Centre for Doctoral Training in Machine Intelligence for Nano-electronic Devices and Systems [EP/S024298/1] and by the Engineering and Physical Sciences
Research Council (EPSRC) ActivATOR project [EP/W017466/1].
The authors acknowledge the use of the IRIDIS High Performance Computing Facility, and associated support services at the University of Southampton, in the completion of this work.
\bibliographystyle{IEEEtran}
\bibliography{bib/IEEEabrv, bib/bibliography_draft}

\vfill

\end{document}